\documentclass[%
 reprint,
 amsmath,amssymb,
 aps,
]{revtex4-2}

\usepackage{graphicx}% Include figure files
\usepackage{orcidlink}
\usepackage{hyperref}
\usepackage{comment}
\usepackage{subcaption}
\usepackage{dcolumn}% Align table columns on decimal point
\usepackage{bm}% bold math
\usepackage[T2A]{fontenc}
\usepackage[utf8]{inputenc}
\usepackage[russian]{babel}
\begin{document}

\preprint{APS/123-QED}

\title{Inelastic collisions facilitating runaway electron generation in weakly-ionized plasmas}

\author{Y. Lee$^1$\orcidlink{0000-0003-4474-416X}}
\author{P. Aleynikov$^2$\orcidlink{0009-0002-3037-3679}}
\author{P.C. de Vries$^3$\orcidlink{0000-0001-7304-5486}}
\author{H.-T. Kim$^4$\orcidlink{0009-0008-2549-5624}}
\author{\\ J. Lee$^5$\orcidlink{0000-0002-2353-2603}}
\author{M. Hoppe$^6$\orcidlink{0000-0003-3994-8977}}
\author{J.-K. Park$^1$\orcidlink{0000-0003-2419-8667}}
\author{G.J. Choi$^1$\orcidlink{0000-0003-0044-1650}}
\author{J. Gwak$^1$\orcidlink{0000-0003-4767-3328}}
\author{Y.-S. Na$^{1}$\thanks{Corresponding author.} \orcidlink{0000-0001-7270-3846}}
\email{ysna@snu.ac.kr}

\affiliation{$^1$ Department of Nuclear Engineering, Seoul National University, Seoul, South Korea}
\affiliation{$^2$ Max-Planck Institute fur Plasmaphysik, Greifswald, Germany}
\affiliation{$^3$ ITER Organization, Route de Vinon sur Verdon, CS 90 046, 13067 St Paul Lez Durance, France}
\affiliation{$^4$ United Kingdom Atomic Energy Authority, Culham Science Centre, Abingdon, OX14 3DB,United Kingdom of Great Britain and Northern Ireland}
\affiliation{$^5$ Korean Institute of Fusion Energy, Daejeon, South Korea}
\affiliation{$^6$ Department of Electrical Engineering, KTH Royal Institute of Technology, SE-11428 Stockholm, Sweden}

\date{\today}

\begin{abstract}
Dreicer generation is one of the main mechanisms of runaway electrons generation, in particular during tokamak startup. In fully ionized plasma it is described as a diffusive flow from the Maxwellian core into high energies under the effect of the electric field. In this work we demonstrate a critical role of the non-differential nature of inelastic collisions in weakly ionized plasma during tokamak startup, where some electrons experience virtually no collisions during acceleration to the critical energy. We show that using the Fokker-Planck collisional operator can underestimate the Dreicer generation rate by several orders of magnitude.
\end{abstract}
\maketitle
\textit{Introduction.}--
Runaway Electrons (REs) are electrons that are accelerated freely by the electric field due to a decreasing drag force with increasing velocity \cite{Wilson1925MPCPS}. They are important not only in astrophysics \cite{Guo2014PRL, Arnold2021PRL} and geophysics \cite{Gurevich1992PLA, Gurevich2013PRL} but also in tokamaks, where RE generation is undesirable due to the danger of triggering plasma instabilities \cite{Snipes2008NF, Liu2023PRL} and ability to damage the device \cite{Nygren1997JNM, Matthews2016PS, Vries2023NF}. In tokamaks REs can be generated during both plasma initiation (startup) \cite{Knoepfel1975PRL, Knoepfel1979NF, Vries2023NF} and disruptions \cite{Lehnen2008PRL, Martin-solis2017NF, Reux2021PRL}. Startup REs are more critical in ITER than in the existing fusion devices because the low plasma density required for plasma burn-through facilitates their generation \cite{Gribov2018EPS, Vries2019NF,Vries2020PPCF, Hoppe2022JPP, Matsuyama2022NF, Lee2023NF} and the large device size ensures their good confinement \cite{Vries2023NF}. Development of a successful startup scenario for ITER and other future reactors thus requires a robust understanding of RE physics and reliable predictions of their generation rates in cold weakly ionised plasmas.

Several RE generation mechanisms are distinguished based on how electrons surpass the critical momentum $p_c$ (force-free) in phase space \cite{Breizman2019NF}. The most important during startup is the Dreicer mechanism, pioneered in Ref. \cite{Dreicer1959PR} and elaborated in subsequent works \cite{Gurevich1961JETP, Lebedev1965JETP, Connor1975NF}. According to this mechanism, a diffusive flux from the Maxwellian core across the critical momentum boundary is formed under the effect of electric field if the electric field exceeds the critical value, $E_c \equiv \frac{e^3 n_e \ln \Lambda}{4 \pi \varepsilon_0^2 mc^2}$, i.e. $E/E_c > 1$. Owing to the dominance of small angle collisions in a fully ionised plasma this process is diffusive and is adequately described by the Fokker-Planck model \cite{Chandrasekhar1943RMP, Rosenbluth1957PR}.

In a weakly ionized plasma, such as a tokamak plasma during startup, electron-atom collisions need to be considered \cite{Townsend1915, Lloyd1991NF, Lloyd1996PPCF, Yoo2018Nat, Chew2024NF}. The effect of the weakly ionized plasma on \textit{highly energetic} electrons may still be appropriately described with the Fokker-Planck operator, since the energy transfer in a typical small angle collision exceeds the ionization potential for highly energetic electrons \cite{Breizman2019NF}. Then, their stopping power is typically described with the Bethe stopping power \cite{Bethe1932}, and elastic scattering can be accounted for with the help of the Thomas–Fermi model \cite{Kirillov1975,Breizman2019NF}. However, in a typical startup the critical momentum may lie in the energy range which is comparable to (or exceeds only by one order of magnitude) the ionization potential, in which case the key role in determining the RE generation rate is played by the region of the phase space for which the small angle approximation is not appropriate. 

Indeed, for instance in ITER startup the inductive electric field is $E \approx0.2 \ V/m$ \cite{Gribov2018EPS, Lee2023PS}, the plasma temperature is increasing, but is expected to be below 10 eV during the plasma breakdown \cite{Yoo2018Nat}, where the free electron density is slowly growing and is initially extremely low. For instance, when $n_e = 10^{17} m^{-3}$, the ratio of the inductive electric field to the Dreicer field, $E_d \equiv \frac{e^3 n_e \ln \Lambda}{4 \pi \varepsilon_0^2 T_e}$, is $E/E_d \approx 0.06$, which is above the known threshold for effective Dreicer generation (0.02-0.025) \cite{Jayakumar1993PRA}, but the corresponding critical momentum, $p_c \approx (E/E_c)^{-\frac{1}{2}}$, is only around 80 eV - a few times higher than the H ionization potential. Evidently, the \textit{non-differential} nature of collisions within this energy range needs to be considered in order to correctly describe electron kinetics.

In this Letter we present numerical calculations of the Dreicer flow accounting for the non-differential nature of inelastic collisions using the Fokker-Planck-Boltzmann operator (FPB) that treats soft and hard collisions with the FP and Boltzmann operator, respectively. The double counting of collisions is avoided by an appropriate selection of the integration boundaries in the Coulomb logarithm definition for the FP operator \cite{Embreus2018JPP}. We demonstrate a significant effect of the non-differential inelastic collisions, which can not be captured using the FP operator.

\textit{Kinetic model.}--
The kinetic equation for the electrons is
\begin{equation}
    \frac{\partial F_e}{\partial t} + eE\Big[ \frac{\partial}{\partial p} \mu +  \frac{\partial}{\partial \mu} \frac{1-\mu^2}{p} \Big] F_e = C^{e,ei} \{ F_e \} + C^{eH} \{ F_e \} \label{eq:kin_eq}
\end{equation}
where $F_e(p,\mu)$ is the distribution function normalized such that the density $n_e = \int F_e(p,\mu) dp d\mu$, $p$ is the particle momentum normalized to $mc$ and $\mu = \cos \theta$ with the pitch angle $\theta$. The FPB operator for collisions between free electrons and hydrogen atoms consists of two parts
\begin{equation}
    \mathcal{C}^{eH} \{ F_e \} = \mathcal{C}^{eH}_{FP} \{ F_e \} + \mathcal{B}^{eH} \{ F_e \} \label{eq:full_kin}
\end{equation}
where $\mathcal{C}^{eH}_{FP} \{ F_e \}$ is the FP operator and $\mathcal{B}^{eH} \{ F_e \}=\mathcal{B}^{eH}_{iz} \{ F_e \} + \mathcal{B}^{eH}_{ex} \{ F_e \}$ is the Boltzman operator. The subscripts $iz$ and $ex$ denote ionization and excitation, respectively.

We introduce a soft-hard separation factor $h$ to determine if an inelastic collision is categorized as soft or hard. When the energy loss fraction of the incident electron is higher than $h$, such collisions are included in the Boltzmann operator; otherwise, they are described by the FP operator. For high $h$ such as $h=0.99$, the test particle operator is approximated as the FP operator. Following the conventional practice, we assume that a scattered electron possesses more energy than an ejected one. The slowing-down frequency for free-bound collisions comprises the soft ionization and excitation collision parts, from $1 \ s \rightarrow 2 \ p$ to $1 \ s \rightarrow 10 \ p$ transitions.
\begin{equation}
    \nu_S^{e,eb} = \frac{\gamma}{p^2} \Big(\nu_{iz}^{sti,soft} + \sum_{n=2}^{10} \nu_{ex}^{1n} E_{1n}' \mathcal{H}(T-\frac{E_{1n}}{h})\Big)
    \label{eq:nuS}
\end{equation}

Here, $\nu_{iz}^{sti,soft}=n_H (\int_0^{W_{bnd}} dW \frac{W+B}{mc^2} \frac{d\sigma_{iz}}{dW}) v$ and $\nu_{ex}^{1n}=n_H \sigma_{ex}^{1s \rightarrow np} v$, where $\frac{d\sigma_{iz}}{dW}$ is the singly differential cross section described by the Relativistic Bineary Encounter Dipole (RBED) model \cite{Kim1994PRA, Kim2000PRA} and $\sigma_{ex}^{1s \rightarrow np}$ is the total excitation cross section with $1s \rightarrow np$ transition by the BE-scaled plane wave Born approximation model \cite{Kim2001PRA, Stone2002JNIST}. %$\beta=v/c$, $\tau_c = \frac{mc}{eE_c}$%$\tau_c = 4\pi \varepsilon_0^2 m^2 c^3 / e^4 n_e \ln \Lambda$
$W_{bnd}=\min \{ \max(0, hT-B), (T-B)/2 \}$, %$n_e$ is the electron density, 
$n_H$ is the hydrogen atom density, $T$ is the incident electron energy, $B$ is the binding energy, $W$ is the secondary electron energy, $E'_{1n}=E_{1n}/mc^2$ is the $1s \rightarrow np$ transition energy $E_{1n}$ normalized to $mc^2$ and $\mathcal{H}$ is the heaviside function. The $W_{bnd}$ is chosen taking into account two conditions: the absence of soft inelastic collisions for $0 \ge hT-B$ and the requirement that the ejected electron energy be less than $(T-B)/2$.

The electron-hydrogen atom soft collisions are collisions against stationary targets. Their FP operator takes the  form \cite{Helander2005Cam}
\begin{equation}
    \mathcal{C}_{FP}^{eH}\{ F_e \} = \frac{\partial}{\partial p}(p \nu_S^{e,eb} F_e) + \frac{\partial}{\partial \mu} \Big[ (\nu_D^{e,eb} + \nu_D^{eH})\frac{\partial F_e}{\partial \mu} \Big].
\end{equation}

There are the deflection frequencies due to collisions with bound electrons and hydrogen atoms, %$\nu_S^{e,eb}=(\gamma^2/p^3)\ln \bar{\Lambda}_{b}^{soft}$, 
$\nu_D^{e,eb}=(1/\gamma) \nu_S^{e,eb}$ and $\nu_D^{eH}=(e^4\gamma/4\pi \varepsilon_0^2 m^2 c^3p^3)(\chi n_H I_2^H(y))$, respectively, where $\gamma$ is the relativistic factor and $I_2^H(y)$ is the second screening coefficient based on the TF theory \cite{Breizman2019NF}, which we fit as $I_2(y) \approx I_2(y_*) + \ln [(y/y_*)^4 + \exp(-I_2(y_*))^4]/4$ with $y=274p$ and $y_*=26$. The TF model is only valid for the energetic particles, we extrapolate the operator to low energies using experimental data \cite{Itikawa1974ADNDT, Buckman2000ECA} and the NIST Standard Database \cite{Jablonski2004JPCRD, Salvat2005CPC}. The correction factor $\chi= 1 + a^{el} (511000 (\gamma-1))^{b^{el}}/(1+c^{el} (511000 (\gamma-1))^{b^{el}})$ is found by fitting, where $a^{el}=79.837201$, $b^{el}=-1.0992754$ and $c^{el}=1.6387662$. %The subscripts $S$ and $D$ designate the normalized slowing down frequency and the deflection frequency, respectively. The superscripts $ee$ and $eH$ distinguish colliding species as bound electrons and hydrogen atoms.
\begin{figure}[t]
    \centering
    \includegraphics[width=\linewidth]{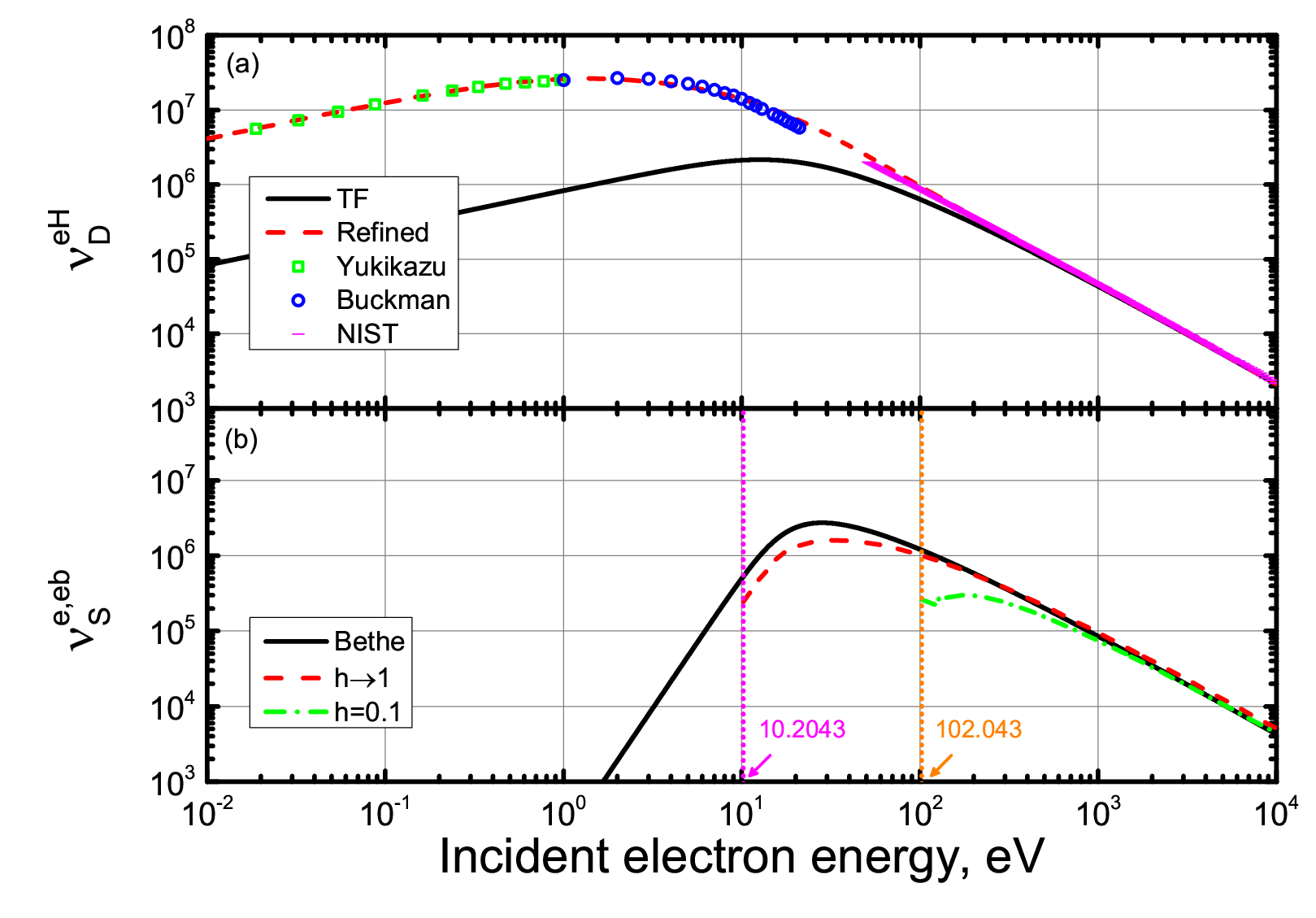}
    \caption{(a) The deflection frequency from the extrapolated TF model (black solid curve), refined model with $\chi$ (red dashed curve), Ref. \cite{Itikawa1974ADNDT} (green square marker), Ref. \cite{Buckman2000ECA} (blue circle marker) and NIST standard database \cite{Jablonski2004JPCRD, Salvat2005CPC} (magenta bar), and (b) slowing-down frequency from the extrapolated Bethe model \cite{Hesslow2017PRL} (black solid curve), the FPB model in $h \rightarrow 1$ limit (red dashed curve) and with $h=0.1$ (green dashed dotted curve) as a function of incident electron energy. $n_e=10^{16} \ m^{-3}$, $n_H=0.99 \cdot 10^{18} \ m^{-3}$ and $E=0.2 \ Vm^{-1}$.}
    \label{fig1}
\end{figure}

Figure \ref{fig1} (a) shows the refined scattering frequency $\nu_D^{eH}$ as a function of the incident electron energy. It is consistent with the TF model at high energy and approximates the experimental data in the low energy region. The soft collisions slowing down frequency $\nu_S^{e,eb}$ (Fig. \ref{fig1} (b)) agrees with the Bethe model for high energy particles. In low energy range, the soft collisions represent only a fraction of all collisions and the frequency is therefore lower. In $h=0.1$ case, for instance, $\nu_S^{e,eb}=0$ and $\mathcal{B}^{eH}$ treats all inelastic collisions below $102.043 \ eV$, this energy corresponds to $1/h$ of the minimal excitation energy for H. We assume that hydrogen is in the ground state. This treatment allows the electrons that didn't experience  inelastic collisions (accounted by $\mathcal{B}^{eH}_{iz}$) not to lose any energy during acceleration to $102.043 \ eV$, the ``continuous'' soft collisions energy loss is significantly reduced even above $102.043 \ eV$. Such accounting of non-differential nature of inelastic collisions changes dramatically the character of critical energy crossing, i.e. it allows {\it some} electrons to accelerate completely freely up to a certain energy.

For $\mathcal{B}^{eH}_{iz}$, we symmetrize the collisional cross section to address the asymmetry in the RBED model originating from its dipole contribution, i.e. $(d\sigma_{iz}/dW)(W,T) = (d\sigma_{iz}/dW)(T-W-B,T) \ if \ W > (T-B)/2$. In addition, a scattering angle distribution is obtained by interpolating the isotropic and Moller scattering distribution \cite{Moller1932AP} as low and high energy asymptotes, respectively, since pitch-angle equilibration is rapid in the low energy region and free-bound collisions can be approximated as free-free collisions in the high energy region. After gyro-averaging,
\begin{equation}
\begin{split}
    \mathcal{B}&^{eH}_{iz} \{ F_e \} = n_H mc^3 \frac{p}{\gamma} \Big[ \int_{p_{bnd}} dp_1 d\mu_1 \frac{d\sigma_{iz}}{dW} \\
    &\times \Big( (\frac{\mathcal{H}((1-\mu^2)(1-\mu_1^2)-(\mu^*-\mu\mu_1)^2)}{\pi\sqrt{(1-\mu^2)(1-\mu_1^2)-(\mu^*-\mu\mu_1)^2}} - \frac{1}{2}) \\
    &\times \frac{1}{1+\frac{0.1}{p}}+\frac{1}{2} \Big)  \frac{p_1}{\gamma_1}F_e(p_1, \mu_1) \Big] - \nu_{iz}^{hard} F_e(p, \mu). \label{eq:iz}
\end{split}
\end{equation}
Here, $\mu^\ast=\sqrt{\frac{(\gamma_1+1)(\gamma-1)}{(\gamma_1-1)(\gamma+1)}}$, $p_{bnd}$ is the momentum of $\min \{ \max(B+W, W/(1-h)), 2W + B \}$, essential for the birth of secondary electron with $p$, $\nu_{iz}^{hard} = n_H ( \int_{W_{bnd}}^{(T-B)/2} dW \frac{d\sigma_{iz}}{dW}) v$ and $1/(1+0.1/p)$ is an interpolation factor.

For $\mathcal{B}^{eH}_{ex}$, we assume the scattering angle to be isotropic:
\begin{equation}
\begin{split}
    &\mathcal{B}^{eH}_{ex} \{ F_e \} = \sum_{n=2}^{10} \Big[- \nu_{ex}^{1n}(p) F_e(p, \mu) \mathcal{H}(\frac{E_{1n}}{h}-T)  \\
    &+  \frac{\nu_{ex}^{1n}(p^+_{1n})}{2} \frac{\beta(p)}{\beta(p^+_{1n})} F_e^0(p^+_{1n}) \mathcal{H}(E_{1n} (\frac{1}{h}-1) - T) \Big] \label{eq:ex}
\end{split}
\end{equation}
where $p^+_{1n} = p(T+E_{1n})$ and $F^0_e = \int d\mu F_e(p, \mu)$.

The full operator $\mathcal{C}^{eH} \{ F_e \}$ retains invariance in particle and energy loss of electrons with respect to $h$, i.e. $\frac{\partial}{\partial h} [ \int dp d\mu \ \mathcal{C}^{eH} \{ F_e \} ] = \frac{\partial}{\partial h} [ \int dp d\mu \ (\gamma-1) \mathcal{C}^{eH} \{ F_e \} ] = 0$:
\begin{align}
    \int dp &d\mu \ \mathcal{C}^{eH} \{ F_e \} = \int dp \Big[ \nu_{iz} F_e^0(p) \Big] \\
    \int dp &d\mu \ (\gamma-1) \mathcal{C}^{eH} \{ F_e \} \notag \\
    &= - \int dp \Big[ (\nu_{iz} B' + \sum_{n=2}^{10} \nu_{ex}^{1n} E_{1n}') F_e^0(p) \Big] \label{eq:invE}
\end{align}
where $\nu_{iz}=n_H (\int_{0}^{(T-B)/2} dW \frac{d\sigma_{iz}}{dW})$ and $B'=B/mc^2$.

The Coulomb collisions with free electrons and ions, $C^{e,ei}$ in Eqn. \ref{eq:kin_eq}, are accounted for using the linearized FP operator as formulated in Ref \cite{Papp2011NF} with the Coulomb logarithms listed in \cite{Breizman2019NF}. This collisional integral assumes that the majority of free electrons are Maxwellian. This can be guaranteed if  $n_e \ln \Lambda_{free} \gg n_H \ln \Lambda_b$, with $\ln \Lambda_{free}$ describing free-free electron collisions and $\ln \Lambda_b$ free-bound collisions. Because $\ln \Lambda_b$ is vanishingly small below  $1s \to 2p$ transition energy, this condition is satisfied and we therefore expect that the lower energy part of the population is predominantly Maxwellian.

\textit{Numerical simulation.}--
We implement the FPB operator, given by Eqn. (\ref{eq:full_kin}), in the self-consistent kinetic simulation \cite{Aleynikov2017NF} using FiPy Finite Volume Partial Differential Equation (PDE) Solver \cite{Guyer2009CSE}.
\begin{figure}[t]
    \centering
    \includegraphics[width=\linewidth]{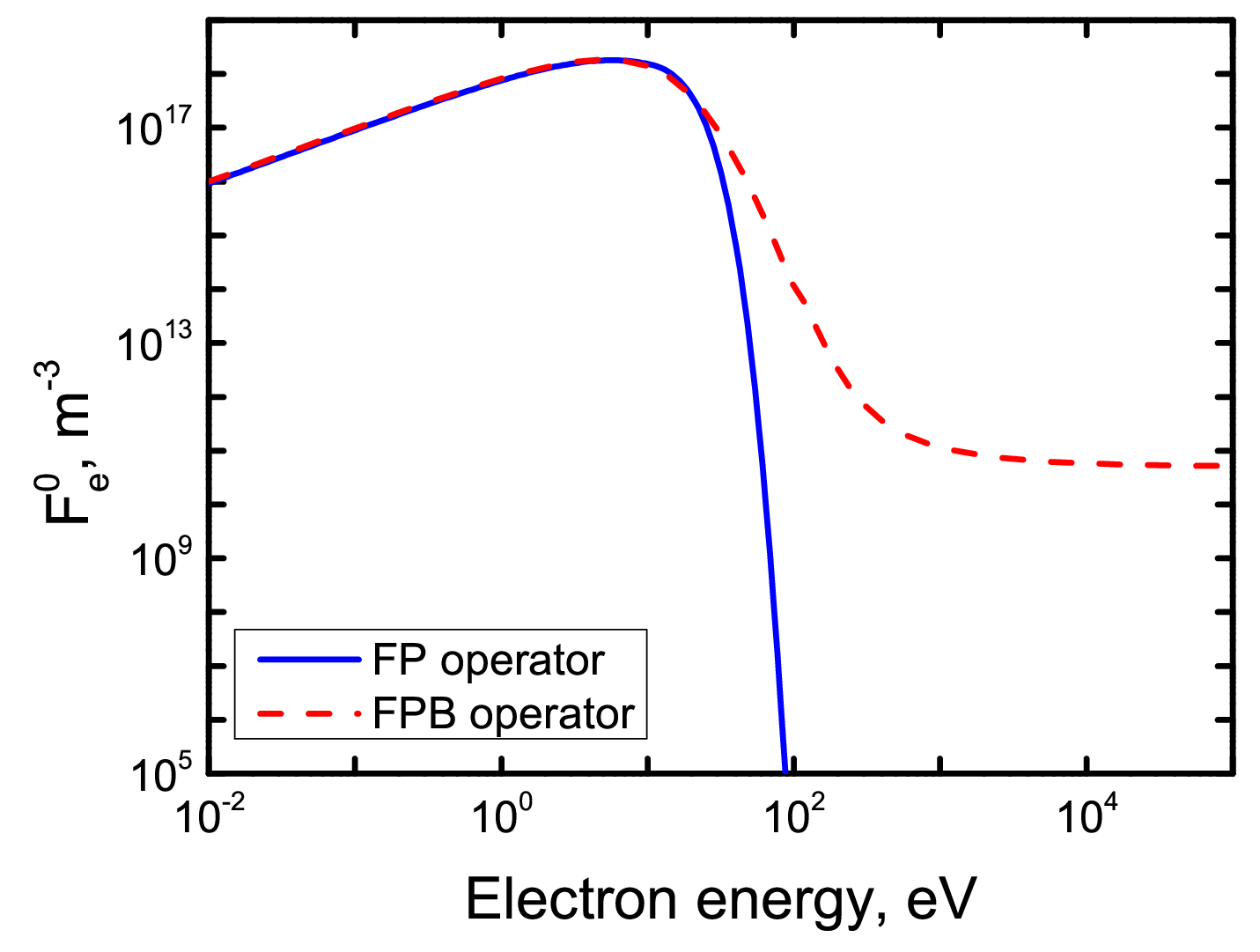}
    \caption{The steady-state solutions of $F_e^0$ with the FPB (red dashed curve, h=0.1) and FP (blue solid curve, h=0.99) operators. Same parameters are used to Fig. \ref{fig1}.}
    \label{fig2}
\end{figure}

The non-differential nature of inelastic collisions facilitates electron acceleration in weakly-ionized plasma. This is demonstrated in Fig. \ref{fig2}, where the steady-state solutions are shown. For the demonstration, we only took the test particle part to eliminate RE avalanche effect \cite{Rosenbluth1997NF} by setting upper boundary of integral in Eqn. (\ref{eq:iz}) to $p$ corresponding to $2W+1$. The blue curve in Fig. \ref{fig2} shows the solution where most collisions are described with the FP operator (h=0.99). The effective $E/E_d^{eff} < 0.02$, with $E_d^{eff} \equiv \frac{e^3 (n_e \ln \Lambda + n_H \ln \Lambda_b)}{4 \pi \varepsilon_0^2 T_e}$ ($\ln\Lambda_{b} \approx 5$) and the resulting $F_e^0 < 10^{3} m^{-3}$ with electron energy above $100 \ eV$. Indeed, no Dreicer flow is expected in such conditions. The red dashed curve shows the solution with the FPB operator ($h=0.1$). This solution is close to Maxwellian in lower energy where free-free collisions dominate. But it differs significantly at higher energy where binary Boltzmann collisions allow for virtually free acceleration for some particles. These freely accelerated particles become runaways. 

\begin{figure}[t]
    \centering
    \includegraphics[width=\linewidth]{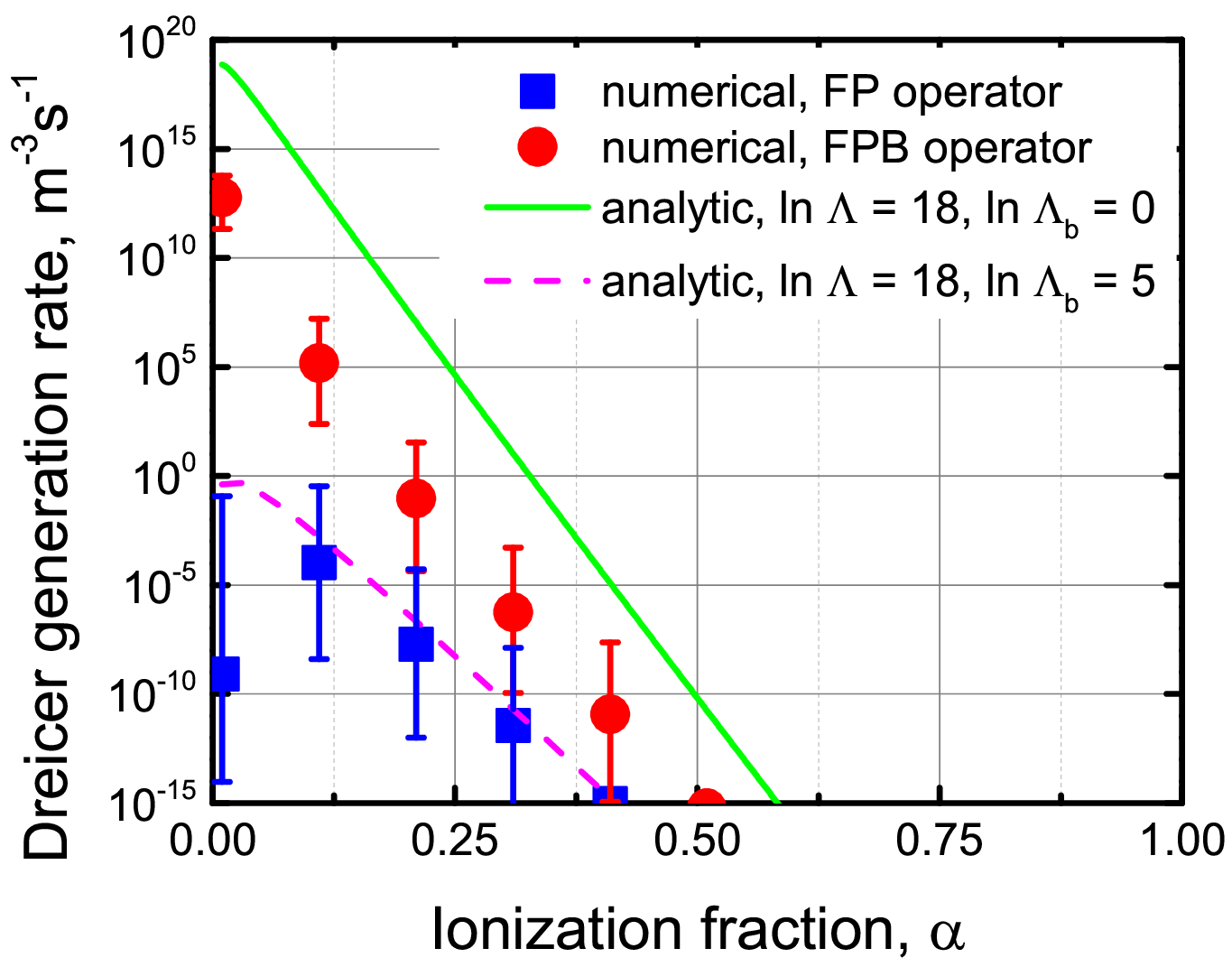}
    \caption{The Dreicer generation rate as a function of $\alpha$: with the FP (blue square, $h=0.99$) and FPB (red circle, $h=0.1$) operators, and the analytic formula \cite{Connor1975NF} with the Coulomb logarithm of free $\ln \Lambda = 18$ and bound electrons $\ln \Lambda_{b} = 0$ (green solid curve) and $\ln \Lambda = 18$ and $\ln \Lambda_{b} = 5$ (magenta dashed curve), required for collisional frequency computation. Here, $(n_H + n_e) = 10^{18} \ m^{-3}$, $T_e  =  5 \ eV$ and $E=0.2 \ Vm^{-1}$. Error bars are estimated by scanning $E=0.18 Vm^{-1}$ to $E=0.22$ ($\pm 10 \%$).}
    \label{fig3}
\end{figure}

Dreicer generation rate as a function of the ionization fraction $\alpha = n_e / (n_e + n_H)$ is shown in Fig. \ref{fig3}. The parameters for this scan correspond roughly to standard Ohmic discharge in KSTAR tokamak \cite{Yoo2018Nat,Lee2023PS} and are similar to that of ITER plasma initiation \cite{Vries2019NF}. In this case, a critical role played by the non-differential nature of collisions is expected for $E/E_d > 0.025$ but $E/E_d^{eff} < 0.02$. The Dreicer generation without accounting for the non-differential nature of collisions is ineffective for these parameters. However, it reaches $10^{13} \ m^{-3} s^{-1}$ for $\alpha = 0.01$ in the case of Boltzmann collisions. This is consistent with the analytical Dreicer rates since $E/E_d \approx 0.43 \gg 0.02$ but $E/E_d^{eff} \approx 0.015$. The observed rates are much lower than the analytic prediction by Connor-Hastie \cite{Connor1975NF} if collisions with bound electrons are ignored in the analytic calculation (green curve in Fig. \ref{fig3}.). We also note that that our calculations using FP operator (blue squares) agree with the Connor-Hastie formula if $E_d^{eff}$ with $\ln \Lambda_{b} \sim 5$ deduced from Eqn.~(\ref{eq:nuS}) is used in the formula (dashed curve).

\begin{figure}[t]
\includegraphics[width=0.5\textwidth]{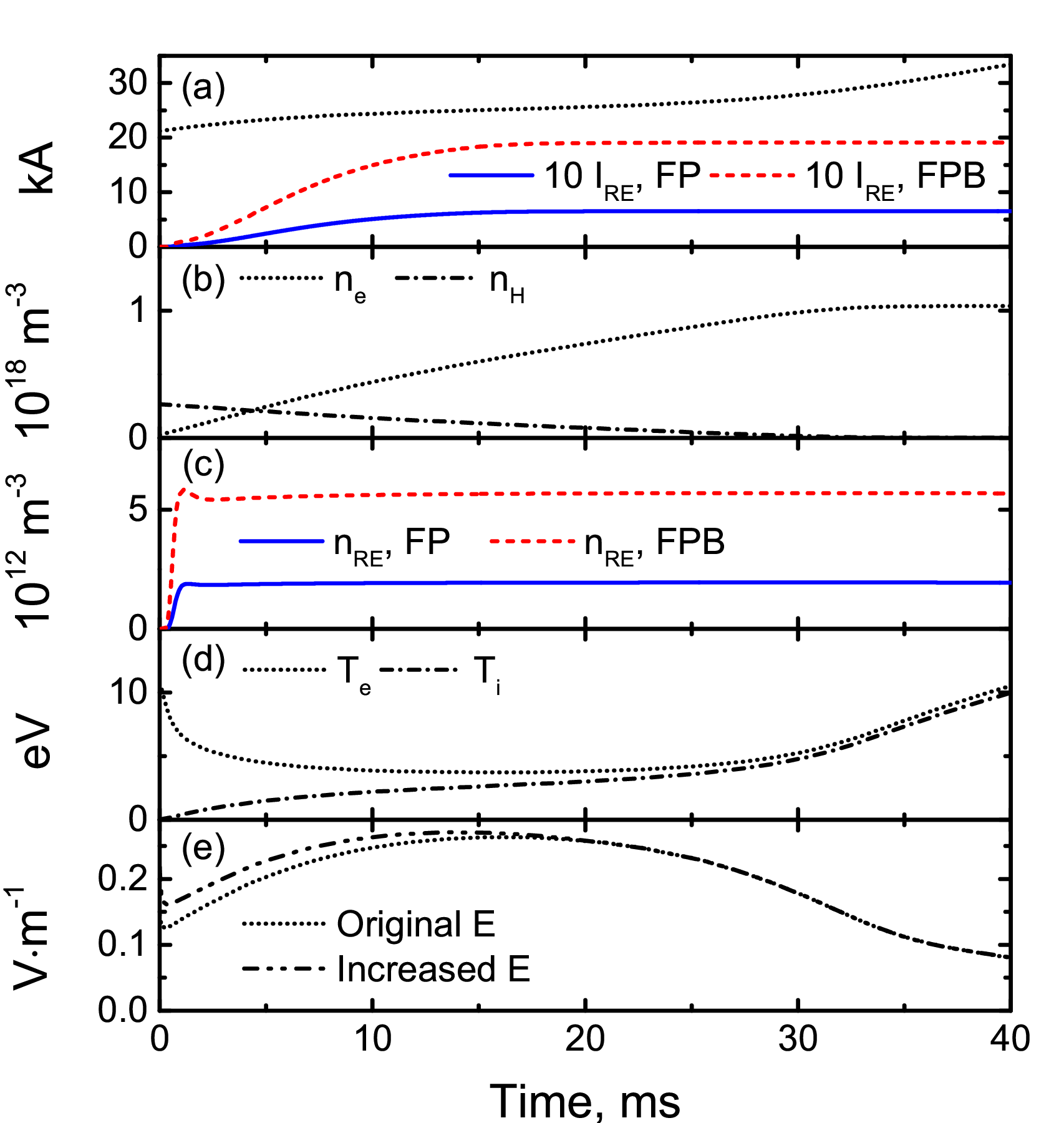}
\caption{$I_{RE}$ (a) $n_{RE}$ (c) during the ITER relevant startup predicted by the kinetic simulation with the FP (red dashed, $h=0.1$) and FPB (blue solid, $h=0.99$) operators, respectively. Plasma parameters are prescribed by DYON (black dotted and dotted-dashed): $I_p$ (a), $n_e$ and $n_H$ (b), $T_e$ and $T_i$ (d), $E$ (e). The increased E is also illustrated (dashed dotted) in (e) to show how higher E results in the complete current takeover. The initial condition is $I_{p} = 21.1 \ kA$ and $\alpha=8 \ \%$ for RE confinement with $V_{loop} = 10.7 \ V$ and $P_{pre}=0.6 \ mPa$.}% DYON is initialized with same $n_e$, $I_{th}$, $T_e$ and $V_{loop}$ (black dotted).}
\label{fig4}
\begin{subfigure}{0pt}
    \phantomcaption
    \label{fig4a}
\end{subfigure}
\begin{subfigure}{0pt}
    \phantomcaption
    \label{fig4b}
\end{subfigure}
\begin{subfigure}{0pt}
    \phantomcaption
    \label{fig4c}
\end{subfigure}
\begin{subfigure}{0pt}
    \phantomcaption
    \label{fig4d}
\end{subfigure}
\begin{subfigure}{0pt}
    \phantomcaption
    \label{fig4e}
\end{subfigure}
\end{figure}

Dreicer generation rate is known to be exponentially sensitive to the value of the electric field \cite{Gurevich1961JETP}. Yet, strong effect of the non-differential nature of collisions should be considered separately: the sensitivity to the electric field is mainly the result of the exponential dependence of the distribution function on $p_c$, wheres in the case of non-differential collisions ($\mathcal{B}^{eH}$) {\it friction-less} acceleration ``through'' $p_c$ of some electrons facilitates the Dreicer flow. The effect of the $\pm 10 \%$ deviation in $E$ is demonstrated with the error bars in Fig. \ref{fig3}.

The Closed Flux Surface Formation (CFSF), before which we assume a complete loss of fast particles, follows the main-speceis burn-through in small-size tokamaks \cite{Yoo2018Nat}. Nevertheless, enhanced Dreicer generation in weakly ionized plasma can have significant implications for the startup scenarios in ITER \cite{Vries2020PPCF} and other current \cite{Vries2019NF, Vries2023NF} and future tokamaks \cite{Bachmann2020FED, Zheng2022Ino, Kim2022SP, Deshpande2023NF} where the flux surfaces may be formed before the complete plasma ionization. 

Figure \ref{fig4} shows kinetic simulation results for the RE density and current during an ITER-relevant startup \cite{Vries2019NF, Vries2023NF}, where plasma parameters $n_e$, $T_e$ and $E$ are prescribed by the ``fluid'' startup code DYON \cite{Kim2012NF, Kim2020NF, Kim2022NF, Lee2023PS}(black curves). $E$ is deduced from the resistive voltage. In this calculation only the phase after the \textit{global} CFSF is simulated, i.e. when the maximum ratio of the stray field to the toroidal field is $10^{-3}$ \cite{Vries2019NF} and therefore the initial plasma current $I_p = 21.1 kA$. The kinetic model considers the full FPB operator in Eqn. (\ref{eq:full_kin}).% and amends the free-free collisions by the knock-on collision operator \cite{Chiu1998NF, Harvey2000POP}. The RE radial transport is modelled by adding $-F_e/\tau_{RE}$ in RHS of Eqn (\ref{eq:kin_eq}) where $\tau_{RE}$ is the Bohm diffusion time \cite{Bohm1949}.

Dreicer generation is most effective early in the simulation, since the rapidly growing electron density $n_e$, which exceeds $n_H$ at $4 \ ms$, makes the generation ineffective quickly as shown in Fig. \ref{fig4b}. Accounting for the non-differential nature of collisions enhances the Deicer generation increasing the ultimate RE density $n_{RE}$ (\ref{fig4c}) and current $I_{RE}$ (\ref{fig4a}) by a factor of 3. In this particular example the resulting $I_{RE}$ overtakes a significant fraction ($7.5 \%$) of the total current, posing a danger of a complete startup failure, due to a reduction of the Ohmic heating (not modeled self-consistently here). Indeed, simulation show that we would already have the complete current takeover if E is only slightly higher as shown in Fig. \ref{fig4e}. This simulation is not shown in Fig. \ref{fig4}, but the ultimate $n_{RE} = 5.6 \cdot 10^{13} m^{-3}$ in this case.

\textit{Conclusions}.--
Reliable startup RE prediction in large-scale future tokamaks such as ITER \cite{Vries2019NF, Vries2023NF} and DEMOs \cite{Bachmann2020FED, Zheng2022Ino, Kim2022SP, Deshpande2023NF} require appropriate description of Dreicer generation during early ionization stage. In this work we have demonstrated an essential role of the non-differential nature of ionizing collision, especially when $E/E_d \geq 0.025$ but $E/E_d^{eff} \leq 0.02$. We show that such enhanced Dreicer generation may lead to a complete failure of tokamak startup if the Ohmic current is taken over by the RE current.

This research was supported by National R\&D Program through the National Research Foundation of Korea (NRF) funded by Ministry of Science and ICT (2021M3F7A1084419). This research was supported by National R\&D Program through the National Research Foundation of Korea (NRF) funded by the Ministry of Science \& ICT (NRF-2019R1A2C1010757). This research was supported by R\&D Program of "High Performance Fusion Simulation R\&D (code No. EN2441-1)" through the Korea Institute of Fusion Energy(KFE) funded by the Government funds, Republic of Korea.
\bibliographystyle{unsrt}
\bibliography{ref}
%\nocite{*}

\end{document}